\documentclass[oldversion]{aa}
\usepackage{txfonts}
\usepackage{graphicx}
\usepackage{natbib}
\bibpunct{(}{)}{;}{a}{}{,}
\begin{document}

\newcommand{\iau}{Int.Astron.U.}
\newcommand{\asj}{Astron.Soc.Jap.}

\title{Dynamics of shock propagation and nucleosynthesis conditions
in O-Ne-Mg core supernovae}
\titlerunning{O-Ne-Mg supernovae and nucleosynthesis conditions}

\author{H.-Th.~Janka, B.~M\"uller, F.S.~Kitaura, and R.~Buras}
\authorrunning{H.-Th.~Janka et.~al.}
\institute{Max-Planck-Institut f\"{u}r Astrophysik,
Karl-Schwarzschild-Str.1, Postfach 1317, 85741 Garching, 
Germany} 
\offprints{Hans-Thomas~Janka, \email{thj@mpa-garching.mpg.de}}
 \institute{Max-Planck-Institut f\"ur Astrophysik, 
              Karl-Schwarzschild-Str.\ 1, D-85741 Garching, Germany
              }
\date{\today}

\abstract{It has been recently proposed that the shocked
surface layers of exploding O-Ne-Mg cores provide the conditions
for r-process nucleosynthesis, because their rapid expansion
and high entropies enable heavy r-process isotopes to form even
in an environment with very low initial neutron excess of the matter.
We show here that the most sophisticated available hydrodynamic
simulations (in spherical and axial symmetry) do not support 
this new r-process scenario because 
they fail to provide the necessary conditions of temperature, 
entropy, and expansion timescale by significant factors.
This suggests that, either the formation of r-process elements
works differently than suggested by Ning et al.\ (2007, NQM07), 
or that some
essential core properties with influence on the explosion dynamics
might be different from those predicted by Nomoto's progenitor model.
}

\maketitle

\section{Introduction}

The site(s) of the production of the r-process elements are still 
a mystery. It has been long speculated that supernova explosions 
of progenitor stars in the $\sim$8$\sim$11$\,M_\odot$ range
play a role in this context, in particular as the
origin of the heaviest r-process nuclei with mass numbers $A > 130$. 
Several arguments have been
brought forward in support of this conjecture. On the one hand,
their progenitors in the mentioned mass window, the most massive of
the so-called super-asymptotic giant branch (super-AGB) stars, 
develop cores that are not made of iron, but of oxygen, neon, and 
magnesium. Since such O-Ne-Mg cores are relatively small, compact,
and bounded by an extremely steep density gradient, their collapse,
triggered by the onset of rapid electron captures, was thought to 
lead to supernova explosions by the prompt hydrodynamical
bounce-shock mechanism. Such explosions have the potential to
eject large amounts of 
highly n-rich (i.e., low electron-to-baryon fraction, $Y_e$) 
matter, in which a strong r-process can happen (Hillebrandt 1978,
Hillebrandt et al.\ 1984,
Sumiyoshi et al.\ 2001, Wanajo et al.\ 2003, Wheeler et al.\ 1998).
On the other hand, considerations of galactic chemical evolution
(e.g., Mathews et al.\ 1992, Ishimaru \& Wanajo 1999, Ishimaru et
al.\ 2005) and observations of metal-poor stars suggest
that the sites of heavy r-process element production are 
decoupled from the main sources of elements between oxygen and
germanium (Qian \& Wasserburg 2002, 2003, 2007). This was interpreted
as support of the speculation that r-nuclei with $A > 130$ 
should be produced in O-Ne-Mg core-collapse supernovae, because
owing to the compact progenitor core these explosions eject very 
little intermediate mass nuclei.

How this production might happen in such supernovae, 
however, is still unclear. The most sophisticated simulations
do not confirm the idea that O-Ne-Mg cores explode by the 
prompt mechanism and thus rule out the possibility of a
low-entropy, low-$Y_e$ r-process in these gravitational collapse
events (Kitaura et al.\ 2006, Mayle \& Wilson 1988;
see also Dessart et al.\ 2006). Stars at the low-mass end of
supernova progenitors are also not the most favorable
sites for strong r-processing in the neutrino-driven wind that
sheds mass off the surface of the hot neutron star 
left behind when the explosion has been
launched. The formation of the third r-process peak in this 
high-entropy, high-$Y_e$ environment was recognized to
require winds from very massive ($M_{\mathrm{ns}}\gtrsim 2\,M_\odot$)
and very compact ($R_{\mathrm{ns}}\lesssim 9\,$km) neutron stars
(Otsuki et al.\ 2000, Thompson et al.\ 2001), which are not
expected to emerge from the collapse of low-mass stars.

Ning et al.\ (2007; NQM07) therefore proposed a new formation
scenario. They argued that heavy nuclei from barium 
through the actinides may be produced in the shocked surface
layers of exploding O-Ne-Mg cores because these layers expand
extremely rapidly after the shock passage, thus allowing high-mass
r-nuclei to be assembled at conditions of moderately large
entropies and $Y_e \sim 0.5$.
Here we will demonstrate that detailed hydrodynamical simulations
of such exploding O-Ne-Mg cores do not yield the conditions that
NQM07 assumed for the expanding shells
from the core surface. This means that either their r-process 
scenario does not take place in O-Ne-Mg supernovae, or the
conditions there are significantly different from current model
predictions.

In Sect.~\ref{sec:models} we will briefly describe the discussed
hydrodynamic explosion models, in Sect.~\ref{sec:results} we
will present our results for the dynamical evolution and explosion
of O-Ne-Mg core supernovae, in Sect.~\ref{sec:nucsyncond} we will
discuss the nucleosynthesis-relevant conditions in the ejecta,
and in Sect.~\ref{sec:conclusions} we will summarize our findings 
and draw conclusions.

\begin{figure}[!t]
\includegraphics[width=8.5cm]{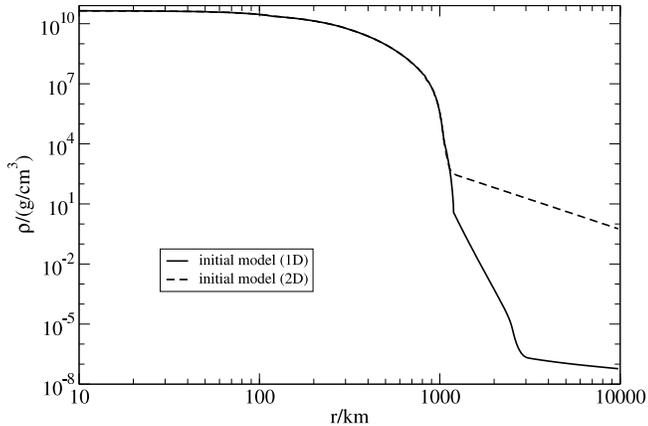}
\caption{Density profiles of the initial O-Ne-Mg core models used for
the 1D and 2D supernova simulations.
The solid curve corresponds to the original core data of
Nomoto (1984, 1987), extended at $\rho \lesssim 10^3\,$g$\,$cm$^{-3}$
by a hydrogen envelope (70\% H, 30\% He) in hydrostatic equilibrium
(Nomoto, private communiation). This stellar structure was used
for the spherically symmetric core-collapse simulation in this
paper. In contrast, the 2D simulation was done with the same core,
but with a dilute, hydrostatic helium shell added at low densities
(dashed line). Such a stellar structure was employed previously by
Kitaura et al.\ (2006).}
\label{fig:density-profiles}
\end{figure}

\begin{figure}[!t]
\includegraphics[width=8.5cm]{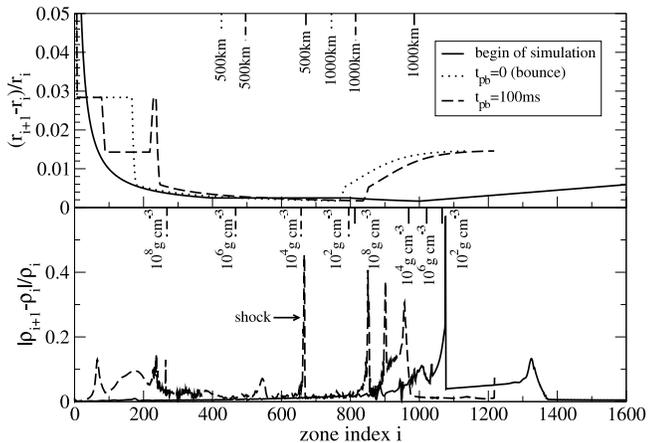}
\caption{
Radial resolution of the 1D simulation at the beginning of
the simulation, at core bounce, and 100$\,$ms after core bounce.
The upper panel shows the relative radius variation, $|\Delta r|/r$,
the lower panel the relative density difference, $\Delta\rho/\rho$,
between neighboring zones as a function of the radial zone index. Note
that the radial grid is comoving with the fluid during collapse and
is kept fix (i.e., Eulerian) after bounce, but then is still refined 
by adding more zones between bounce and 100$\,$ms later (for reasons
of better visibility
only two times are given in the lower plot). In the panels the 
positions of selected radii and densities are marked by vertical
bars in the same line styles as the curves.
One can see that the radial spacing is better than 0.3\% and 
the density change less than 10\% in the region of the steep 
density gradient between about 100$\,$g$\,$cm$^{-3}$ and 
$10^7\,$g$\,$cm$^{-3}$.
}
\label{fig:radial-resolution}
\end{figure}

\section{Computed models}
\label{sec:models}

\begin{figure}[!t]
\begin{center}
\includegraphics[width=8.0cm]{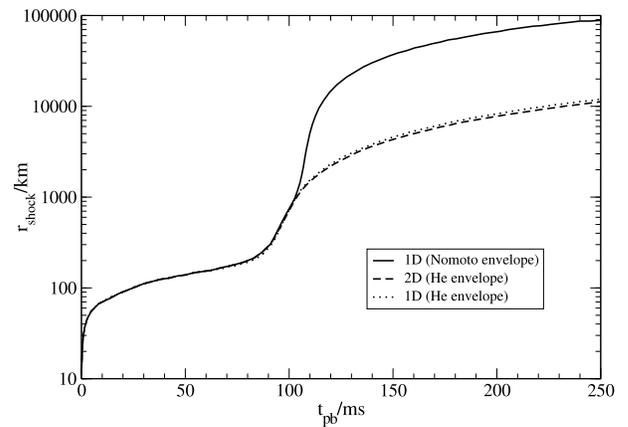}
\end{center}
\caption{Radii of the supernova shock as functions of time
for the 1D and 2D simulations (solid and
dashed lines, respectively). The stronger acceleration of the shock
in the region outside of about 1100$\,$km is a consequence of the
steeper density decline of the model employed in the 1D run
(Fig.~\ref{fig:density-profiles}). While the 2D simulation was done
with the 8.8$\,M_\odot$ progenitor with artificially constructed
low-density He-shell at $\rho \lesssim 10^3\,$g$\,$cm$^{-3}$ (see
Kitaura et al.\ 2006), the 1D simulation was performed with a
recently updated progenitor structure, in which a much more
dilute H-envelope was added around the O-Ne-Mg core
(K.~Nomoto, private communication). The shock trajectory for 
a 1D run with the same progenitor structure as in the 
2D model is also plotted for comparison (dotted line) and is nearly
indistinguishable from the dashed curve.}
\label{fig:shock-radii}
\end{figure}

We discuss here results of core-collapse and explosion
simulations for an 8.8$\,M_\odot$ star with an 1.3776$\,M_\odot$
O-Ne-Mg core
(Nomoto 1984, 1987). One was conducted in spherical symmetry
(1D) with the initial density profile given by the solid line in
Fig.~\ref{fig:density-profiles}. Another model was two-dimensional
(axisymmetric; 2D) and was computed with a less steep density decline
below $\rho \sim 10^3\,$g$\,$cm$^{-3}$, represented by the
dashed line in Fig.~\ref{fig:density-profiles}. (For reasons of
comparison, a 1D run was also performed with the progenitor
profile of the 2D simulation and the shock trajectory of this
calculation will be shown in Fig.~\ref{fig:shock-radii}.)

The reason for the use of two different
density structures outside of the O-Ne-Mg core is historical.
The initially available data file of the 8.8$\,M_\odot$ star
only contained data above a density of 
$1.44\times 10^3\,$g$\,$cm$^{-3}$,
but no information was given for the stellar layers at radii
$r > 1.095\times 10^8\,$cm. Kitaura et al.\ (2006)
therefore extended the model with a dilute shell of helium in
hydrostatic equilibrium, being guided by the structure above
the iron core of slightly more massive progenitors.
More recently, Nomoto (private communication) provided a data
table in which a hydrostatic hydrogen envelope was added around
the thin $\sim$0.1$\,M_\odot$ carbon-oxygen shell 
(between $\sim$4$\times 10^8\,$g$\,$cm$^{-3}$
and $\sim$3$\times 10^4\,$g$\,$cm$^{-3}$) and the even thinner
shell of $\sim$4$\times$10$^{-6}\,M_\odot$ of helium 
(between $\sim$3$\times 10^4\,$g$\,$cm$^{-3}$ and
$\sim$6$\times 10^3\,$g$\,$cm$^{-3}$). The structural
difference of the two initial density profiles plotted in
Fig.~\ref{fig:density-profiles} has no influence on the onset of
the supernova explosion and the energy of the explosion. It also
plays no role for the nucleosynthesis conditions in the density
regime between $\sim$10$^8\,$g$\,$cm$^{-3}$ and
$\sim$10$^3\,$g$\,$cm$^{-3}$, which is the matter of discussion
in this paper. But of course, it has influence on the subsequent
propagation and acceleration of the outgoing supernova shock.

Both simulations were performed with the Lattimer \& Swesty (1991)
equation of state (EoS) at high densities. 
Kitaura et al.\ (2006) conducted a 1D run also for the
Hillebrandt \& Wolff EoS (Hillebrandt et al.\ 1984), 
which is stiffer around and above
nuclear matter density than the Lattimer \& Swesty (1991) EoS.
The outcome of the simulations for both EoSs was qualitatively
the same and even quantitatively extremely similar with respect
to the shock formation and propagation, the mass cut, and the
explosion properties. Also the other elements of the input physics
were the same as in Kitaura et al.\ (2006), except
for the use of an updated version of the electron capture rates
on nuclei in nuclear statistical equilibrium (NSE), which were
improved compared to the previous version of Langanke et al.\ (2003)
by adding electron screening corrections and a more refined 
description of the neutrino emission spectrum 
(K.~Langanke, G.~Mart\'{\i}nez-Pinedo,
and J.M.~Sampaio, private communication). Another (smaller) difference
with minor consequences for the dynamical evolution was the inclusion 
of inelastic neutrino scattering off nuclei in 
NSE as described by Langanke et al.\ (2008).

For the simulations discussed here we employed non-equidistant, 
time-dependent radial grids. In the hydrodynamics module 
of our code we used 1600 Lagrangian zones during collapse and 
between 1150 (within the first 80$\,$ms p.b.) and 1720 Eulerian zones 
after core bounce. The neutrino transport was done with
221 to 411 radial cells; 
coarser grid spacing than for the hydrodynamics was chosen
in the (nearly) transparent layers where the
neutrino-matter interactions become irrelevant. Moreover, the
outer boundary of the transport grid after bounce was put to 
2000$\,$km
instead of the $10^5$--$10^7\,$km of the hydrodynamics grid.
In setting up
the latter, particular care was taken of a high resolution in the steep
density gradient at the surface of the O-Ne-Mg core. 
Figure~\ref{fig:radial-resolution} shows the radial resolution as 
a function of the zone index in terms of the relative density and radius
differences between neighboring zones, $|\Delta\rho|/\rho$ and 
$\Delta r/r$, respectively, at three (two) representative times:
at the start of the 1D simulation, 100$\,$ms after core bounce, and
in one case also at the moment of bounce.
One can see that in the steep density
gradient at the core surface the density varies from zone to zone 
typically by less than ten percent and the radius by less than 
0.3 percent.
The 2D model had 128 lateral zones of the polar grid. 

For doing the simulations of O-Ne-Mg core collapse presented
here and in Kitaura et al.\ (2006), the implementation of nuclear
burning and of electron captures was significantly modified
and extended
compared to the code description given in Rampp \& Janka
(2002) and Buras et al.\ (2006). A simplified treatment 
accounts now for the main thermonuclear reactions involving seven
symmetric nuclei (He, C, O, Ne, Mg, Si, Ni). Their abundance
changes are described by computing successively the analytic, 
time-dependent solutions of the rate equations of two- or 
three-particle reactions, beginning with the fastest of the
included reactions (details will be given in a forthcoming
paper by M\"uller et al.\ 2008). The energy released by 
the nuclear burning in the non-NSE regime is carried effectively
away by electron captures (see, e.g., Miyaji \& Nomoto 1987,
and references therein), of which those on $^{20}$Ne and
$^{24}$Mg are the most important ones for the considered ensemble
of nuclei. The corresponding rates were taken from 
Takahara et al.~(1989). 

We point out that our description of the thermonuclear
energy production without a full reaction network is 
approximative and it might be desirable to improve on that in 
future simulations, also including electron capture rates in a large
network fully consistently. However, in combination with our present
treatment of electron captures, our simplified implementation
of nuclear burning is sufficiently accurate to ensure a smooth,
essentially transient-free transition from the progenitor evolution 
of Nomoto's model to the collapse phase computed with our code. 
Initially pressure and gravity forces keep the core very 
close to hydrostatic equilibrium, and heating by nuclear reactions 
is nearly balanced by cooling through neutrino emission; ongoing
contraction of the central core regions is a consequence of a
slight bias towards neutrino losses (see Kitaura et al.\ 2006 for
a discussion of this critical point). For these reasons we think
that our approach is more than adequate to describe the contraction
of the O-Ne-Mg core during the very early stages of the infall. 
The C+O shell at the surface of the core, whose radial 
structure is most relevant for the
discussions of the present paper, begins to collapse only
when the pressure support from the deeper layers breaks down because
an increasingly larger inner part of the core gets involved in the
collapse. With the rising temperature nuclear burning of carbon 
and oxygen in this non-NSE region accelerates dramatically, 
but the burning
timescale does not come close to the dynamical timescale before 
the infall velocities have become
supersonic. The transition from fuel to ashes of different 
burning stages then occurs in rather narrow radial regions.
The nuclear energy release there leads to a transient 
deceleration of the still highly supersonic infall, which shows
up as sawtooth-like features on the velocity profile.
It is possible that a more sophisticated treatment of nuclear
burning and electron captures affects the details of this 
behavior, but we do not see a reason why one should expect 
that a more refined
network description might lead to a fundamentally different 
dynamical behavior of the supersonically infalling shells.

\begin{figure}[!t]
\begin{center}
\includegraphics[width=8.5cm]{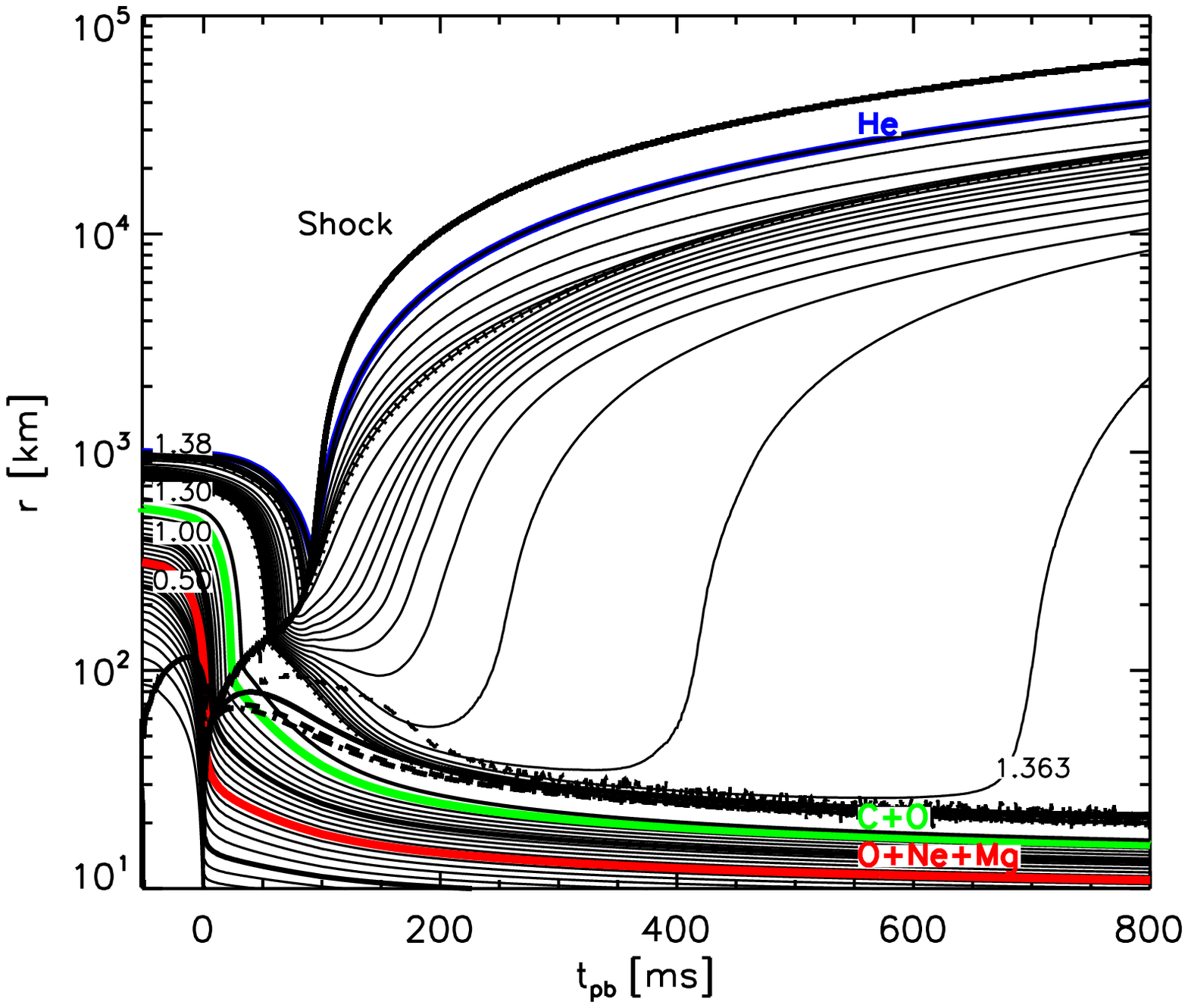}
\end{center}
\caption{Mass shell trajectories for the 1D simulation with the 
helium envelope model as functions of post-bounce time $t_\mathrm{pb}$
(cf.\ Fig.~\ref{fig:shock-radii} for the shock trajectory
of this progenitor and Fig.~1 in Kitaura et al.\ (2006) for an
explosion simulation of the same model with the stiffer EoS of 
Hillebrandt et al.\ 1984). 
Also plotted are the shock position (bold solid
line starting at time zero and rising to the upper right corner),
the gain radius (thin dashed line), and the neutrinospheres for
$\nu_e$ (thick solid line), $\bar{\nu}_e$ (thick
dashed line), and $\nu_\mu$, $\bar{\nu}_\mu$, $\nu_\tau$, 
$\bar{\nu}_\tau$ (thick dash-dotted line). In addition, the composition
interfaces of the progenitor core are plotted with different bold 
colored lines: the inner boundary of the O-Ne-Mg layer at 
$\sim$0.72$\,M_\odot$ (red), of the C-O layer at $\sim$1.23$\,M_\odot$
(green), and of the He layer at 1.3775$\,M_\odot$ (blue). The two dotted
lines represent the mass shells where the mass spacing between the 
plotted trajectories changes. An equidistant spacing of 
$5\times 10^{-2} M_\odot$ was chosen up to 1.3579$\,M_\odot$, 
between that value and 1.3765$\,M_\odot$ it was $1.3\times 10^{-3}
M_\odot$, and $8\times 10^{-5} M_\odot$ farther outside. {\em (A color
figure is available in the online version of our paper.)} }
\label{fig:massshells} 
\end{figure}

\section{Dynamical evolution and explosion}
\label{sec:results}

\begin{figure*}[!t]
\includegraphics[width=17.5cm]{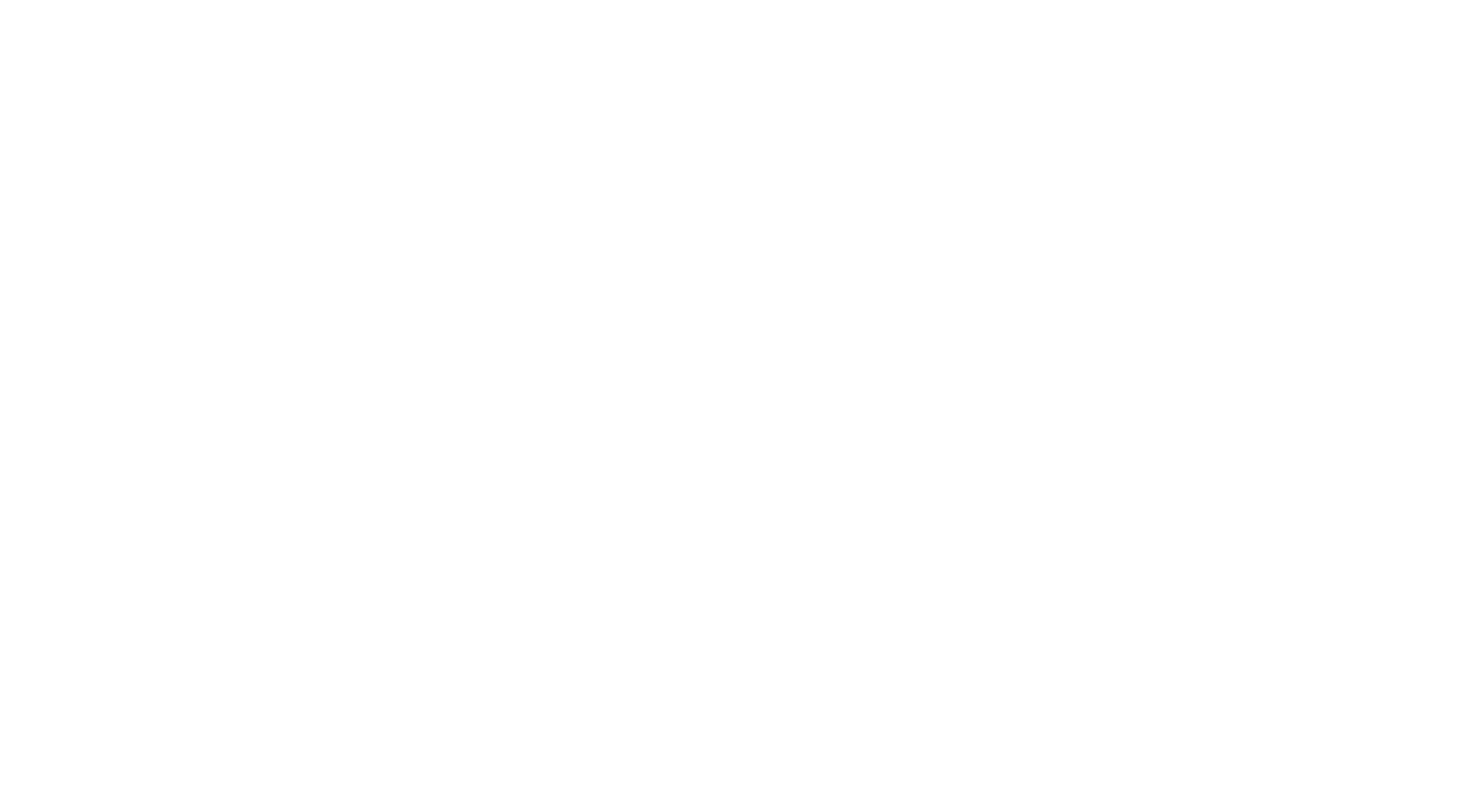}
\caption{
{\em Panel~a:} Explosion energies of the 1D and 2D runs
as functions of time after bounce. The plotted energy
is the sum of thermal plus degeneracy, kinetic,
and gravitational energies, integrated for all matter
with a positive value of this quantity at a certain time (see
Eqs.~27 and 29 in Buras et al.\ 2006).
The 2D simulation becomes slightly more energetic due to the
effects of convective overturn in the gain layer.
{\em Panels~b, c, d:} Time evolution of energies that
characterize the energy budget of different ejecta shells or regions
in the 1D model with hydrogen envelope. {\em Panel~b} shows the
energies for a mass shell close to the O-Ne-Mg core
surface (between mass coordinates 1.376913$\,M_\odot$ and
1.3769486$\,M_\odot$) with the lower boundary chosen at the
location of the shell for which NQM07 performed their
nucleosynthesis studies (see also Fig.~\ref{fig:trho-history}).
{\em Panel~c} corresponds to the mass outside of an enclosed mass
of 1.3675$\,M_\odot$; this mass shell is blown out at the time
when the explosion energy in panel~a makes the steep rise (the
inner boundary of this shell is located at about 150$\,$km at
200$\,$ms after bounce). {\em Panel~d} shows the integrated
quantities for all matter outside of a mass coordinate
of 1.3626$\,M_\odot$, which defines the preliminary mass cut
at the end of our simulation (see Fig.~\ref{fig:massshells}).
The black line denotes the evolution of the total energy
(i.e., the internal energy plus renormalized rest-mass energy
contribution plus kinetic energy minus gravitational binding energy), 
the red line gives the volume and time integral of the
net energy deposition by neutrinos in the gain region, the green
curve the integrated net energy loss in the neutrino-cooling region,
the blue curve the cumulative compressional ($P\mathrm{d}V$)
work exchanged between the considered mass shell and the settling or
expanding neutron star, and the orange curve the sum of these three
effects, which follows well the behavior of the total energy.
The dashed magenta line visualizes the cumulative
energy that is converted between internal and rest-mass energy
through nuclear burning, nuclear photodisintegration, and
nucleon recombination (see text for a detailed explanation).
{\em (Color
figures are available in the online version of our paper.)}
}
\label{fig:explosion-energy}
\end{figure*}

\begin{figure}[!t]
\includegraphics[width=8.5cm]{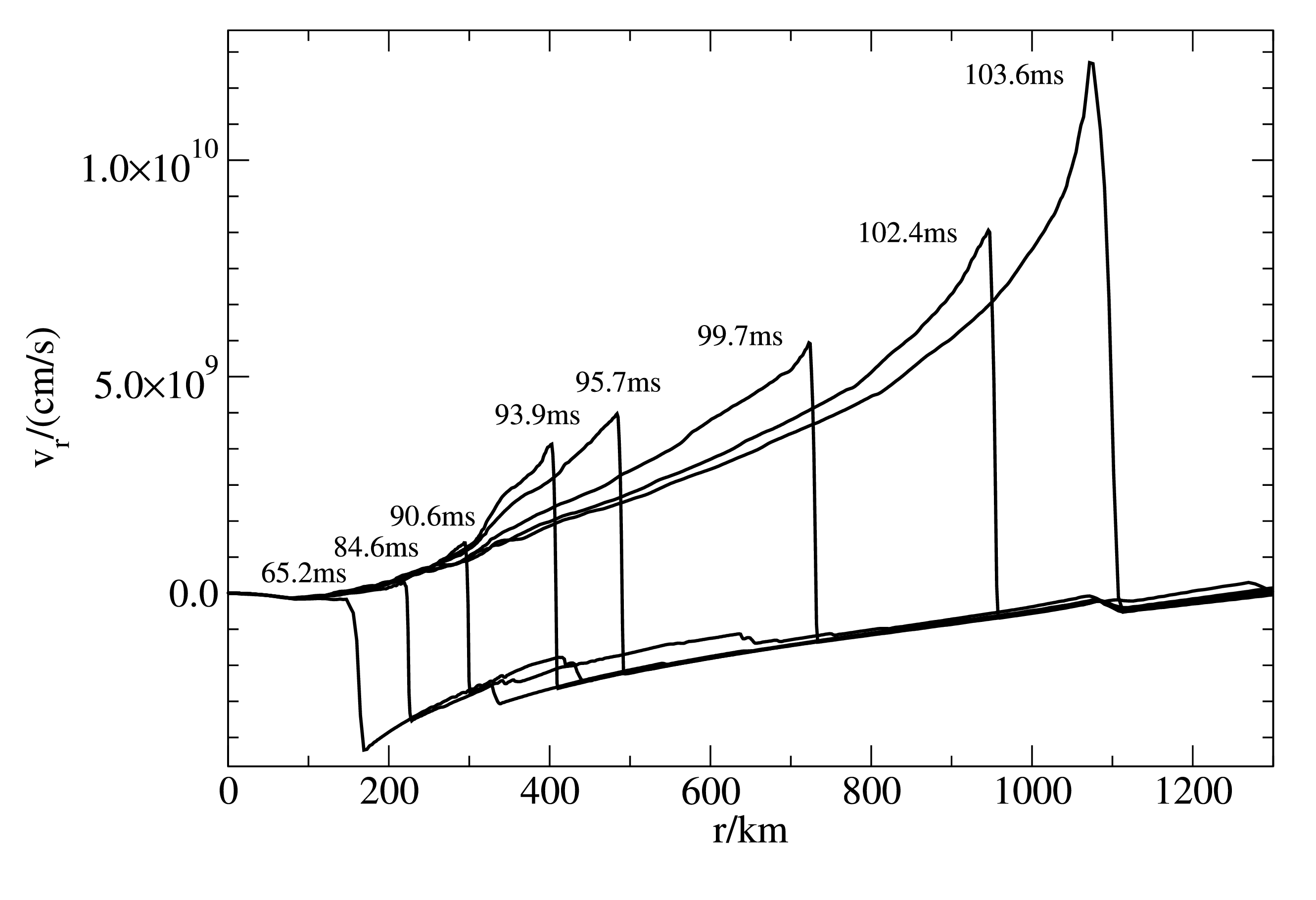}
\caption{Radial velocity profiles from the 1D simulation for different
post-bounce times as indicated in the plot. The shock accelerates 
as it propagates down the steep density gradient at the surface of 
the O-Ne-Mg core, reaching velocities of more than one third of the 
speed of light approximately 103$\,$ms after bounce. }
\label{fig:velocity-profiles}
\end{figure}

Supernovae of low-mass progenitors like the considered 8.8$\,M_\odot$
star with O-Ne-Mg core can be powered and driven by the 
neutrino-heating mechanism (Kitaura et al.\ 2006; Mayle \& Wilson 1988).
The explosions of the two 1D and 2D simulations discussed here 
develop in the same way as described in detail by Kitaura et al.\ (2006).
The shock radii as functions of time are displayed in 
Fig.~\ref{fig:shock-radii}. 
The difference between the two shock trajectories 
is entirely caused by the 
different density profiles shown in Fig.~\ref{fig:density-profiles}.
A comparison of 1D and 2D runs with exactly the same progenitor
structure confirms that it is {\em not} the result of multi-dimensional
physics being ignored in the one case but playing a role in the other
(see Fig.~\ref{fig:shock-radii}).
The reason for this insensitivity of the early shock propagation
to the dimensionality of the 
simulation is the fact that the shock on its way out of the 
O-Ne-Mg core accelerates enormously when it runs down the steep
density gradient bounding the core (Fig.~\ref{fig:velocity-profiles}). 
Its evolution is essentially unaffected by the 
convective overturn that develops in the neutrino-heated
layer just above the gain radius ($R_\mathrm{gain}\sim 100\,$km at
$t\sim 100\,$ms p.b.\ and $R_\mathrm{gain}< 50\,$km at 
$t\ga 200\,$ms p.b., see
Fig.~\ref{fig:massshells}), because convective overturn in this
region becomes strong only later than $\sim$100$\,$ms after bounce (see
Fig.~2 in Janka et al.\ 2007 and Fig.~1 in Janka et al.\ 2008).
At this time the shock already crosses a radius of 1000$\,$km 
(Figs.~\ref{fig:shock-radii} and \ref{fig:velocity-profiles})
and is therefore far away from the convective layer just outside
of the gain radius. Since the shock
propagates with high velocity to large distances, the sound
crossing time from the convective layer to the shock grows so
quickly that sonic communication cannot take place on the simulated 
timescales. Therefore the developing convective activity around the
neutron star has no effect on the shock and the shock trajectory 
does not reveal differences between 1D and 2D simulations. In
contrast, convection in the gain layer and neutron star has 
moderate consequences for the explosion energy of the supernova,
which becomes about $10^{50}\,$erg at the end of our simulations 
(Fig.~\ref{fig:explosion-energy}, panel~a).

The explosion energy in panel~a of Fig.~\ref{fig:explosion-energy}
at a certain time is defined as the sum of thermal plus degeneracy
energy (i.e., internal energy without rest-mass energy), kinetic 
energy, and gravitational energy of all matter where this sum, which 
we call ``local binding energy'', is positive at the given time
(cf.\ Eqs.~27 and 29 in
Buras et al.~2006, however in the present work evaluated with the 
effective relativistic potential of Case~A in Marek et al.~2006,
which was also used in our simulations). One should note that the
mass that fulfills the ``explosion criterion'' (i.e., positive local
binding energy) varies with time. For the considered progenitor
star with its loosely bound hydrogen envelope, which does not yield
a significant additional energy contribution, the final value of
the explosion energy thus
defined is equivalent to the excess energy of the supernova ejecta at
infinity. A steep rise in the explosion energy occurs between 140$\,$ms
and 260$\,$ms after core bounce shortly after the first mass in the 
neutrino-heating layer has begun to expand outward from locations 
close to the gain radius
(see Fig.~\ref{fig:massshells}). This steep rise is mainly caused
by a very rapid increase of the mass that has obtained
positive local binding energy, which means that more and more mass
shells fulfill the explosion criterion. It is
at this time that the matter initially forming the gain layer 
becomes gravitationally unbound. 
Even slightly before (at about 100$\,$ms
after bounce) the explosion energy had reached a little plateau of 
some $10^{48}\,$erg. This plateau is associated with a small amount of 
material that was swept outward when the shock accelerated in rushing
down the steep density gradient at the surface of the O-Ne-Mg 
core. The positive energy of this matter was transferred by 
$P{\mathrm d}V$ work from the expanding and pushing outer
layers of the nascent neutron star just below the ejected mass shells
(see panel~b of Fig.~\ref{fig:explosion-energy}, which 
will be further discussed in the next paragraph).
Even earlier (at $t_\mathrm{pb}\approx 60\,$ms) the gain radius
had developed and neutrino heating had started to deposit energy in 
the postshock layer (see Fig.~\ref{fig:massshells} and panels~c and d
of Fig.~\ref{fig:explosion-energy}). The time delay between this moment
and the onset of the steep rise of the explosion energy is caused by
the fact that the matter in the newly established gain layer is
gravitationally bound and neutrino heating has to deposit enough energy
before the local binding energy of this gas can become positive.
After roughly 260$\,$ms p.b., the rise of the explosion energy flattens.
At that time the gas that was initially in the gain layer has expanded
outward and the gain radius has retreated to the neutron star surface.
Subsequently, more mass is continuously ablated (with a decreasing
rate) from the surface of 
the nascent neutron star in the neutrino-driven wind, whose power is
responsible for the gradual increase of the explosion energy over
longer timescales (see also Woosley \& Baron 1992, 
Qian \& Woosley 1996, Thompson et al.\ 2001, Arcones et al.\ 2007
and references therein).

In panels~b--d of Fig.~\ref{fig:explosion-energy} we 
display the time evolution of different energies that account for the 
energy budget of selected ejected mass shells or mass regions
(the corresponding layers are defined
in the figure caption) in the 1D model. 
The red line gives the volume and time integrated 
net energy deposition by neutrinos in the gain region, the green
curve the integrated net energy loss in the neutrino-cooling region,
the blue curve the cumulative compression ($P\mathrm{d}V$) work exerted
on the settling neutron star or transferred to the considered mass shell 
by its expanding surface layers, and the orange curve is the sum of
these three effects, which follows well the behavior of the
total energy as represented by the black line. The latter displays the
time evolution of the ``total energy''. In contrast to the local
binding energy integrated for the explosion energy in panel~a
of Fig.~\ref{fig:explosion-energy}, this total energy is defined as the 
internal plus kinetic minus gravitational binding energy {\em plus a 
rest-mass energy contribution}\footnote{Instead of just the internal 
(i.e.\ thermal plus degeneracy) energy, the total energy contains the
relativistic energy of the nucleons, i.e.\ their rest-mass energies plus
their internal energy, renormalized by subtracting 930.773$\,$MeV per 
nucleon. The latter roughly corresponds to the rest mass of nucleons
bound in iron-group nuclei.}, which ensures that
nuclear photodisintegration and recombination effects do not show 
up in the time evolution of the total energy. This makes sense because
these effects do not yield any significant
net contribution to the energy balance
of a collapsing and subsequently ejected mass shell, nor do they 
contribute to the excess energy (i.e., explosion energy) of the 
supernova ejecta. The latter fact can be immediately verified by 
inspecting the dashed magenta line, which displays the cumulative
energy that is exchanged between internal and rest-mass energy
through nuclear composition changes by burning (very small 
positive contribution),
photodisintegration (responsible for a negative derivative), 
and nucleon recombination (leading to a positive derivative).
Converting the plotted total energy to the total binding energy as
volume integral of the local binding energy requires adding the
values of the black and magenta lines.

For each mass shell or region the time evolution of the total energy
and its cumulative energy gains and losses can be visualized in 
such a plot, providing insight into the effects
that determine its dynamics and decide about its approach to a
gravitationally unbound state and its 
contribution to the supernova energy. 
In Fig.~\ref{fig:explosion-energy} such valuable information
is given for three exemplary cases. In panel~c all mass outside
of a mass coordinate of 1.3675$\,M_\odot$ is considered; this
shell is representative of the phase when the steep rise of the
explosion energy in panel~a occurs (its inner boundary is located
at about 150$\,$km at 200$\,$ms after bounce). 
In panel~d the integration
includes all mass above the mass cut that develops until 
the end of the
simulation (the inner boundary of this shell is associated 
with a mass coordinate of 1.3626$\,M_\odot$, compare 
Fig.~\ref{fig:massshells}), and in panel~b
the evaluated ejecta layer is enclosed by the mass
coordinates of 1.376913$\,M_\odot$ and 1.3769486$\,M_\odot$.
The latter shell corresponds to the mass associated with the
surface region of the O-Ne-Mg core (chosen such that the ejecta 
shell considered by NQM07 for their nucleosynthesis studies is
included; see also Fig.~\ref{fig:trho-history}) but it accounts 
only for a small fraction of the core matter that gets ejected
in the explosion. This shell becomes unbound immediately after
it is hit by the shock
(at about 90$\,$ms after bounce) and obtains its positive energy
of about $10^{48}\,$erg by the $P$d$V$ work of the expanding 
deeper layers (see the blue line in panel~b of 
Fig.~\ref{fig:explosion-energy}, which accounts for the growth 
of the total energy after shock passage);
this and the adjacent mass
shells at the O-Ne-Mg core surface produce the small plateau
before the steep rise of the explosion energy visible in panel~a
of Fig.~\ref{fig:explosion-energy} (cf.\ discussion above). 

For the dominant part of the ejecta that come from the O-Ne-Mg 
core (panels~c and d of Fig.~\ref{fig:explosion-energy}),
neutrino heating in the gain layer (red curve) provides by far 
most of the energy that the shells finally contribute to the
explosion energy (black curves at the end of the simulated 
post-bounce period) and compensates for the energy losses due to 
compression work on the neutron star interior (blue line) and due to
neutrino emission at times when parts of the layer are inside
the cooling region
(green line). The black and orange curves in panel~d of
Fig.~\ref{fig:explosion-energy} show the time-integrated
evolution of the total energy of all ejected O-Ne-Mg core mass
from the beginning until the end of our 1D simulation: the shells
start out from a marginally bound state in the progenitor core 
(with a total energy of roughly
$-4\times 10^{49}\,$erg), then first lose energy by $P$d$V$ work
during the beginning collapse (until about 60$\,$ms after bounce),
then receive energy by neutrino heating after the gain radius has 
formed at $t\ga 60\,$ms p.b. (see Fig.~\ref{fig:massshells}),
but transiently can again (panel~d) or not (panel~c) 
lose more energy by compression work to the 
forming neutron star at times when the latter shrinks rapidly 
(until about 300$\,$ms after bounce) before finally the 
contraction of the inner shell boundary slows down sufficiently
that neutrino heating in the considered shell becomes clearly
dominant. Only afterwards the total energy of the integrated ejecta 
mass (panel~d) rises continuously and in fact steeply, because 
neutrinos deliver the energy that lifts the matter from its
gravitationally bound state to an unbound state with excess
energy. This, of course, happens later for mass
shells that get blown out later, corresponding to their inner 
boundary being deeper inside the neutron star.
At the end of our simulation the black
and orange lines in panel~d of Fig.~\ref{fig:explosion-energy}
match the temporary value of the explosion energy plotted in 
panel~a\footnote{A very small remaining difference stems from 
the rest-mass contributions that are per definition included in 
the total energy but not in the explosion energy at a time when
the recombination of nucleons and $\alpha$-particles to nuclei in 
the ejecta is still incomplete.}.
 
Let us now discuss in more general terms the physical processes
that play a role for the development of O-Ne-Mg core explosions
and for providing their power.
The onset of the explosion of stars with O-Ne-Mg core is 
facilitated by the very steep density gradient at the edge of the 
core. This rapid density decline allows the shock
to expand in response to the rapidly decreasing mass accretion
rate and the associated drop of the ram pressure of infalling
material (this was already discussed in detail by Kitaura et al.\ 2006).
We emphasize that this outward acceleration of the shock at the
time when the steep surface gradient reaches it, and the thus 
triggered reexpansion of the postshock gas, cannot be the
cause of the supernova explosion associated with the positive
ejecta energy visible in Fig.~\ref{fig:explosion-energy}.
This energy has to be provided by some sufficiently strong 
source, for which different possibilities exist in our 
(nonrotating and nonmagnetic) models:
(i) $P{\mathrm d}V$ work from also expanding but not ejected
layers at the lower boundary of the region of outward mass 
acceleration; the associated energy is found to be very small
(at most some $10^{48}\,$erg, see panels~b--d of 
Fig.~\ref{fig:explosion-energy} and the corresponding discussion 
in the text); (ii) energy
release by the recombination of free nucleons to $\alpha$-particles
and heavier nuclei; (iii) nuclear burning in matter swept up by
the outgoing shock; (iv) neutrino energy deposition, and 
(v) a flux of sonic energy associated with sound waves originating
from a violently turbulent accretion layer at the neutron star
surface or/and from large-amplitude g-mode oscillations of the
neutron star core. This has recently been suggested to play
a crucial role in the acoustic explosion mechanism
(Burrows et al.\ 2006, 2007), but its relevance is controversial 
(Weinberg \& Quataert 2008).
A quantitative evaluation reveals a negligible contribution
from this effect to the ejecta energy in the discussed models
(details will be presented in M\"uller et al.\ 2008).

How important are the other potential sources of energy, (ii)--(iv),
when we ask for the origin of the excess energy of the bulk of the
ejecta?
First one should note that the gas behind the shock and close to 
the neutron star is strongly bound in the gravitational potential
of the forming compact remnant (the internal energy plus kinetic
energy minus gravitational binding energy of a nucleon at 100$\,$km 
is typically $\la -15\,$MeV at $t \ga 100\,$ms after bounce).
The matter in the infall region ahead
of the shock starts out as part of the progenitor star from a 
gravitationally bound state (about $-2\,$MeV per nucleon at the
outer edge of the O-Ne-Mg core). 
The local binding energy of this matter becomes even
more negative when it goes through the shock and the shock 
heating causes the photo-disintegration of nuclei to
free neutrons, protons, and $\alpha$ particles, a process in which
several MeV per nucleon are converted from thermal energy to 
nucleon rest-mass energy. A sizable amount of energy is also
lost through $P\mathrm{d}V$ work exerted by the infalling mass
shells on the settling neutron star (additional
neutrino energy losses can occur but are only relevant for material 
that gets accreted to locations below the gain radius; see
Fig.~\ref{fig:explosion-energy}, panel~d, in comparison to panel~c,
where this is not the case). As a consequence
of all these energy losses in neutrinos, nuclear dissociation, and
compressional work transferred to the neutron star, the matter in
the gain layer and the surface of the neutron star is much more
bound than it was before its collapse. The recombination
of nucleons during a possible later re-expansion and ejection of
this matter --- point (ii) of the list above ---
can at most return the energy consumed earlier by nuclear
dissociation and can thus help lifting the matter back towards a state
near marginal gravitational binding, which it had before it was
accreted and photodisintegrated in the shock or behind the
shock. This is obvious from the dashed magenta line in
panel~d of Fig.~\ref{fig:explosion-energy}, which 
starts at zero and comes back there at the end.
Nucleon recombination releases energy that was temporarily
stored in rest-mass energy and converts it back to thermal
energy. This energy release raises the pressure and unquestionably
can thus have a very important
influence on the dynamics of the supernova gas. In particular, it
can assist and support the shock expansion and outward  
acceleration of the matter behind the shock, because it happens
right at the time when the gas starts to cool as it begins to move
away from the neutron star in reaction to the energy input by 
neutrinos. Nuclear recombination energy by itself, however, cannot
bring the gas energy to significant positive values\footnote{This
assessment is based on considering the {\em effective net energy balance}
of some collapsing and ultimately ejected matter, which means that 
the initial, gravitationally bound state of the
gas (composed of heavy nuclei) in the core of the progenitor star 
is compared with the final state of the gas after ejection. 
Our conclusions are valid independent of the exact moment and 
detailed reason of the nuclear photodisintegration, whether such
dissociation happens as a consequence of the
compressional heating during infall, due to shock heating, or
because matter is bathed in the intense neutrino flux of the nascent
neutron star. A small net
gain of energy can in principle be obtained only when the recombination
leads to more strongly bound nuclei than the undissociated matter
started out from. This could account for at most $\sim$$10^{49}\,$erg
per $10^{-2} M_\odot$ of ejected matter if the pre-collapse material
consisted for example of oxygen and carbon while the ejecta contained 
mostly iron-group nuclei (see the dashed magenta line in 
panel~c of Fig.~\ref{fig:explosion-energy}). 
Such a gain of energy could occur either through nuclear
burning or less directly by photodisintegration and later recombination 
when the matter goes through NSE.}. This can be
achieved only by the processes mentioned in items (iii) and (iv).

Thermonuclear burning in the shock-heated matter, source (iii), 
which might play a role for the explosion of more massive progenitor
stars (Mezzacappa et al.\ 2007), contributes to the blast
energy of O-Ne-Mg core supernovae only on a minor level.
A firm upper limit of the thermonuclear energy production
can be estimated from the 
fact that $\la 10^{-2}\,M_\odot$ of nickel are ejected (and at most
an order of magnitude less oxygen), corresponding to 
$\la$10$^{49}\,$erg of energy from nuclear burning (less than
1$\,$MeV per nucleon when oxygen or silicon are converted to iron). 
This, however, largely overestimates this contribution, because by 
far most of the ejected iron-group material originates from matter
that was very hot and in NSE before it got ejected and began cooling.
Such material is already included in the energy budget by item (ii).
The only remaining power source for explaining the growing
positive explosion energy is therefore neutrino heating in the 
gain layer. Panel~d of Figure~\ref{fig:explosion-energy}
displays the time-integrated energy that is transferred
by neutrinos to the ejected matter in the gain layer. This
contribution can well account for the energetics of the developing 
explosion; in fact it is much larger since neutrino energy deposition 
also helps to bring the ejecta out of their gravitationally bound
state close to the neutron star. Because of convective
overturn in the gain layer, which carries cool gas to radii near
the region of strongest neutrino heating, the 2D simulation 
accumulates slightly more power than the 1D model, although 
convection has no influence on the propagation of the shock in 
O-Ne-Mg core supernovae.

\begin{figure}[!t]
\begin{center}
\includegraphics[width=8.5cm]{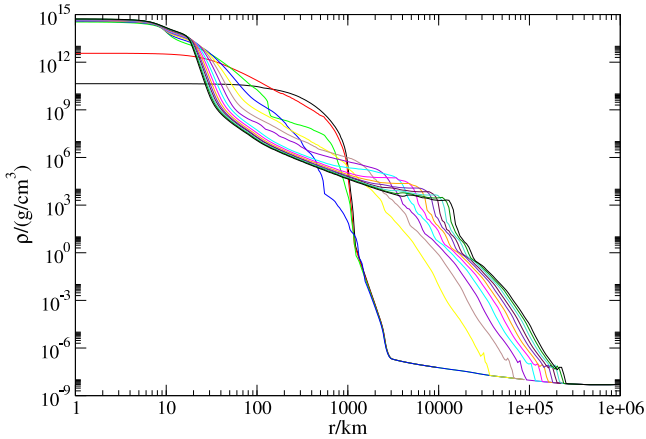}
\end{center}
\caption{Density profiles at
$t=0,\,50,\,100,....,\,700\,$ms
after the start of the 1D simulation (core bounce is at $t = 53.6\,$ms).
At $t = 100\,$ms (46.4$\,$ms p.b.) the supernova shock
is visible at $r \approx 130\,$km, at $t = 150\,$ms (96.4$\,$ms p.b.)
it is near 500$\,$km (see also Fig.~\ref{fig:velocity-profiles}), and
for $t \geq 200\,$ms its position coincides
with the lower right footpoints of the density slopes.
Note the significant flattening of the initially steep density gradient
at the core surface due to the partial and differential collapse of
these layers before shock passage. {\em (A color
figure is available in the online version of our paper.)} }
\label{fig:density-evolution}
\end{figure}

\section{Nucleosynthesis-relevant conditions}
\label{sec:nucsyncond} 

Having in mind the extremely fast expansion of the shocked surface
layers of O-Ne-Mg cores,
NQM07 advocated an r-process scenario for such rapidly
expanding matter. In this case the neutron-to-proton ratio can be 
close to unity (Meyer 2002), provided the entropy is sufficiently high,
around $s\sim 150\,$k$_{\mathrm{B}}$ per nucleon. NQM07 assumed
that such entropy values are produced by the outgoing shock in the 
carbon-oxygen shell around densities of  $\rho \sim 10^6\,$g$\,$cm$^{-3}$,
where still enough matter is located to allow for the production of an
interesting amount of r-process material. They, moreover, assumed 
that the gas is heated by the shock to NSE temperatures 
($T \sim 10^{10}\,$K) before it starts expansion with a timescale
of order 1$\,$ms. For this to be achieved, the shock was considered
to propagate with a velocity of 
$v_{\mathrm{sh}} = 1.5\times 10^{10}\,$cm$\,$s$^{-1}$.

NQM07 used the shock-jump relations to connect
preshock and postshock conditions (density, velocity, and pressure)
and employed
an analytic approach to describe the evolution of the shock-accelerated
mass shells. To this end they made the simplifying assumption of
a strong shock, zero preshock velocity, adiabatic expansion,
and relativistic gas particles (radiation and electron-positron
pairs). In addition, they had to employ an assumption for the 
shock velocity and its dependence on the preshock density,
in which case they could derive expressions for the density $\rho(t)$
and the temperature $T(t)$ of the shocked, expanding mass elements as 
functions of time $t$. Moreover, they considered the shock running
with its assumed speed through the unmodified progenitor core 
structure. This is only a crude approximation. 
In reality, the core has started to contract before the shock 
reaches its surface layers. Since deeper regions of the core begin
to collapse first, the absolute value of the infall velocity develops 
a maximum at the edge of the homologously collapsing inner core and
decreases towards larger radii at any given time. Therefore the 
accelerating contraction proceeds in a differential way.
This leads to a significant flattening of the initially very
steep density gradient around the C/O shell before the shock
hits this region (Fig.~\ref{fig:density-evolution}).

In contrast to the approximative treatment by NQM07, 
we determine the dynamics
and thermodynamics of the supernova gas from our sophisticated
numerical explosion models. In the following, we will compare the
nucleosynthesis relevant conditions in the supernova ejecta as
obtained in the simulations with those assumed by NQM07.

Analytically, using the Rankine-Hugoniot shock-jump relations,
the postshock temperature $T_{\mathrm{p}}$ and entropy 
$s_{\mathrm{p}}$ (in units of Boltzmann's constant
$k_{\mathrm{B}}$ per nucleon) can be written as functions of 
the preshock conditions in the following way
(see Eqs.~2 and 3 in NQM07):
\begin{eqnarray}
T_{\mathrm{p}} &\approx& 1.05\times 10^{10}\,\rho_{\mathrm{pre},6}^{1/4}
\left ( v_{\mathrm{sh},10} - v_{\mathrm{pre},10}\right )^{1/2}\,
\mathrm{K}\,, \label{eq:temperature} \\
s_{\mathrm{p}} &\approx& 56.1 \frac{(v_{\mathrm{sh},10}-
v_{\mathrm{pre},10})^{3/2}}{\rho_{\mathrm{pre},6}^{1/4}}\ \
k_{\mathrm{B}}\ \mathrm{nucleon}^{-1}\,, \label{eq:entropy} 
\end{eqnarray}
where $v_{\mathrm{sh}}$ is the velocity of the shock and 
$v_{\mathrm{pre}}$ and $\rho_{\mathrm{pre}}$ are the velocity
and density, respectively, of the gas ahead of the shock. 
The velocities are normalized to $10^{10}\,$cm$\,$s$^{-1}$,
and the density to $10^6\,$g$\,$cm$^{-3}$. In contrast to 
NQM07, we have considered here the more general
expressions for the case in which the preshock gas is not at
rest.

\begin{figure}[!t]
\includegraphics[width=8.5cm]{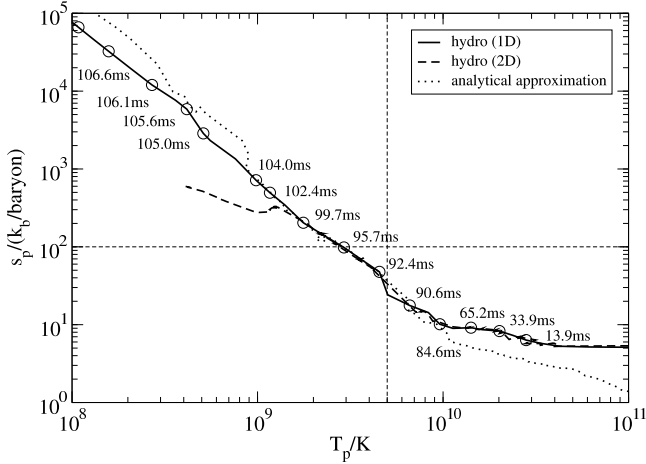}
\caption{Postshock entropy $s_{\mathrm{p}}$
versus postshock temperature $T_{\mathrm{p}}$ for the
1D and 2D hydrodynamic simulations (solid and dashed line, respectively).
The labeled open circles correspond to the post-bounce times 
when the shock hits the mass shells in the 1D run
(as in Fig.~\ref{fig:velocity-profiles}). Note that the dashed curve
begins to deviate from the solid one at $T_{\mathrm{p}}\lesssim 10^9\,$K
when the shock in the 2D calculation slows down at propagating into
the more shallow density profile of the He-shell added outside of the
O-Ne-Mg core (see dashed line in Fig.~\ref{fig:density-profiles}),
while in the 1D simulation the shock continues to accelerate.
The dotted curve depicts the case when the postshock temperature
and entropy are computed from Eqs.~(\ref{eq:temperature}) and
(\ref{eq:entropy}), respectively, using the results of the 1D hydrodynamical
model for the density and velocities on the rhs of these equations.
The conditions in entropy-temperature space considered by
NQM07 for r-processing are roughly above the
horizontal short-dashed line and right
of the vertical short-dashed line. The latter approximately
marks the boundary of the region where the shock produces temperatures
that allow the shocked gas to reach nuclear statistical equilibrium.
}
\label{fig:entropy-temperature}
\end{figure}

\begin{figure}[!t]
\includegraphics[width=8.5cm]{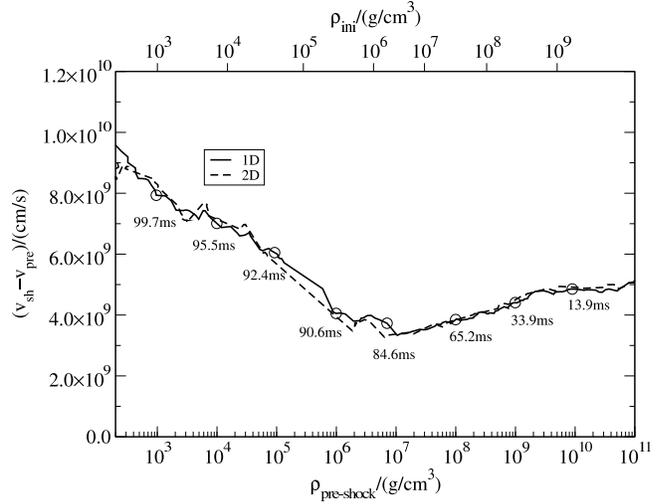}
\caption{Velocity of the shock, $v_{\mathrm{sh}}$, relative to
the preshock gas, whose velocity is $v_{\mathrm{pre}}$, versus
preshock density as obtained from the 1D (solid line) and 2D
(dashed line) simulations. Post-bounce times of the 1D run are
indicated by the labeled open circles (as in
Figs.~\ref{fig:velocity-profiles} and \ref{fig:entropy-temperature}).
For orientation, the scale on the upper horizontal axis 
approximately gives the initial densities of the progenitor model.
These are smaller than the values on the lower horizontal axis
for layers which have started to collapse before they
are reached by the outgoing shock.
}
\label{fig:vstoss-density}
\end{figure}

\begin{figure}[!t]
\includegraphics[width=8.5cm]{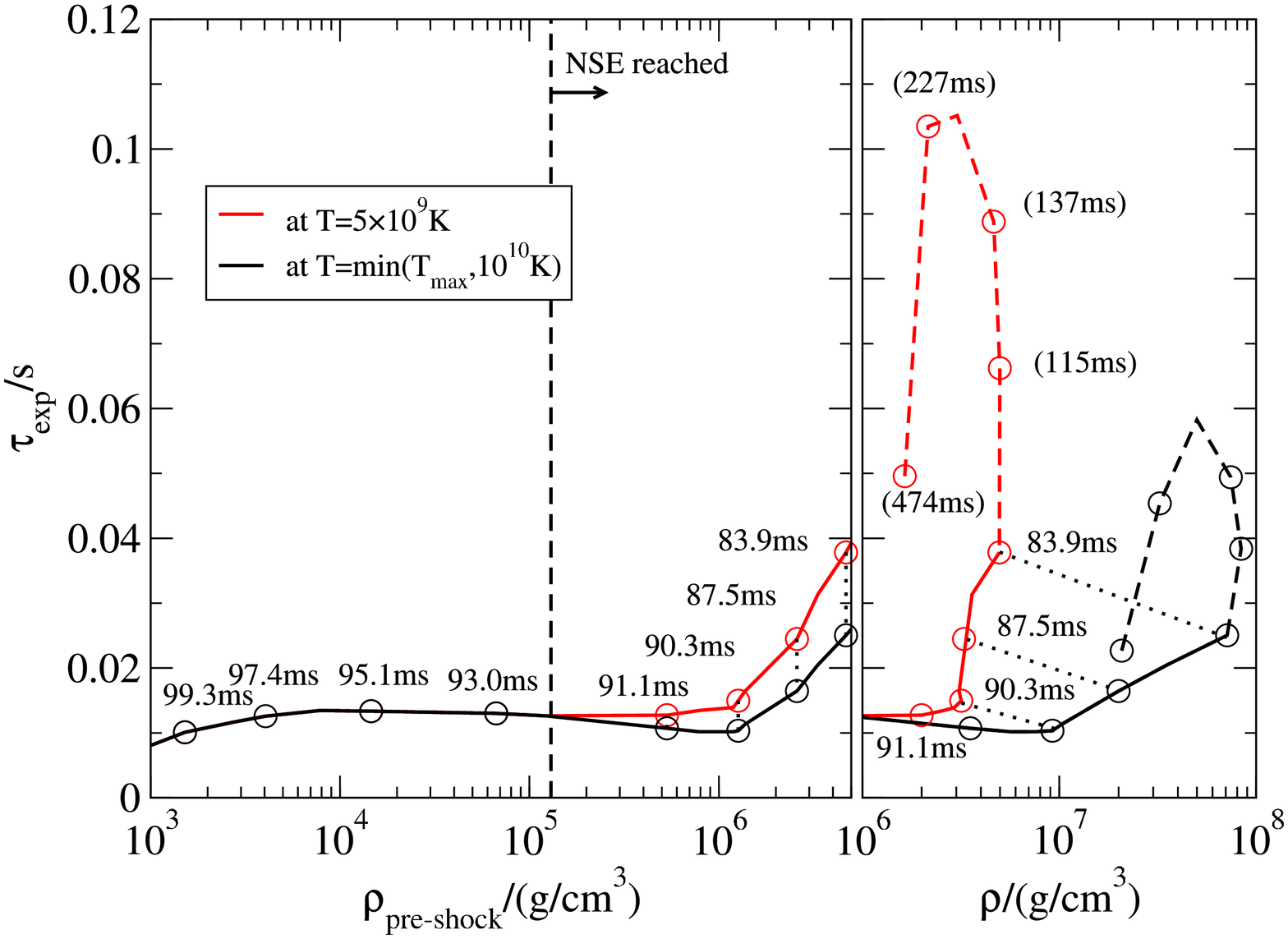}
\caption{Expansion timescales $\tau_{\mathrm{exp}}$ of the
supernova ejecta versus density for the 1D
simulation. The red curve shows the time it takes
the gas in an ejected mass shell to cool
from $T = 5\times 10^9\,$K to $T/e$,
The black curve provides the cooling time between
the peak temperature reached in an ejected shell (or at most
$10^{10}\,$K) at the indicated post-bounce instants (open
circles) and $1/e$ of this value. Points on both curves
that belong to the same mass shell can be identified by
the same post-bounce moments (open circles connected
by thin dotted lines). Only for densities
$\rho \gtrsim 10^5\,$g$\,$cm$^{-3}$ is the
shocked gas heated to conditions near nuclear statistical
equilibrium (i.e., $T \gtrsim 5\times 10^9\,$K, marked
by the vertical dashed line). The solid curve
represents gas that is accelerated outward immediately when it
is hit by the outgoing shock (at the indicated moments),
whereas the dashed curves correspond
to matter that is first accreted onto the nascent
neutron star to be later ablated again from the neutron star
surface in the neutrino-driven wind. Note that in the left panel
the horizontal axis gives the preshock density, while in the
right panel it shows the density that corresponds to the reference
temperature for the timescale measurement. {\em (A color
figure is available in the online version of our paper.)}
}
\label{fig:expansion-time}
\end{figure}

In Fig.~\ref{fig:entropy-temperature} we show the postshock entropy
vs.\ the postshock temperature as computed in the hydrodynamic
models. The 1D and 2D results are in perfect agreement until the
shock in the 2D case reaches a density of about 
$10^3\,$g$\,$cm$^{-3}$, where the progenitor density profiles 
of the two simulations begin to differ (Fig.~\ref{fig:density-profiles}).
This is the case at $t \ga 100\,$ms after bounce as can be seen from
the time labels in Figs.~\ref{fig:velocity-profiles} and 
\ref{fig:density-evolution}. The dotted line in
Fig.~\ref{fig:entropy-temperature} represents the analytic behavior
obtained from Eqs.~(\ref{eq:temperature}) and (\ref{eq:entropy}).
For the velocities and the preshock gas density on the rhs of these
equations we used the values from the 1D simulation. In the temperature
window between $T_{\mathrm{p}} \approx 10^9\,$K and 
$T_{\mathrm{p}} \approx 10^{10}\,$K, which is the relevant one for
our present considerations, the analytic values and the numerical
results are in good agreement. Only in the regimes of lower and higher
temperatures, some of the assumptions made in the derivation of 
Eqs.~(\ref{eq:temperature}) and (\ref{eq:entropy}) are not well
fulfilled any more and the agreements becomes worse.

We note, however, that the entropy-temperature combinations
produced by the shock are much different from those needed for
the r-process scenario considered by NQM07. 
In regions where the shock heats the matter to temperatures
where NSE can be reached (at least $T_{\mathrm{p}} = 5\times 10^9\,$K),
the entropies stay below $s_{\mathrm{p}} \approx 30\,$k$_{\mathrm{B}}$
per nucleon, while the temperature remains less than 
$\sim$2$\times 10^9\,$K in those layers where the postshock entropies 
become around or larger than $s_{\mathrm{p}} \sim 150\,$k$_{\mathrm{B}}$
per nucleon.
Nowhere the temperature and density of the shocked gas simultaneously
reach the conditions desired by NQM07, which are roughly 
in the region above the horizontal short-dashed line and to the right of
the vertical short-dashed line in Fig.~\ref{fig:entropy-temperature}.

The reason for this failure is clear from 
Fig.~\ref{fig:vstoss-density}. While NQM07 assumed a shock
velocity of $v_{\mathrm{sh}} = 1.5\times 10^{10}\,$cm$\,$s$^{-1}$, 
the actual shock speed in the hydrodynamic models (more precisely,
the shock speed relative to the preshock gas) is always less than
$8\times 10^9\,$cm$\,$s$^{-1}$ for $\rho > 10^3\,$g$\,$cm$^{-3}$
and even less than $6\times 10^9\,$cm$\,$s$^{-1}$ for 
$\rho > 10^5\,$g$\,$cm$^{-3}$ (Fig.~\ref{fig:vstoss-density}).

The slower shock also leads to much longer expansion timescales
of the shock-accelerated shells than considered by NQM07.
We define the expansion timescale $\tau_{\mathrm{exp}}$
of mass shells ejected in the
supernova explosion by the time it takes the gas to cool from
a temperature $T$ to $1/e$ of this value. This timescale can
be related to the times $\tau_1$ and $\tau_2$ used by 
NQM07 to characterize the expansion of the
surface area of a mass element and the increase of its 
thickness, respectively, by the following relation:
\begin{equation}
(1 + x)^2 \left ( 1 + x\, \frac{\tau_1}{\tau_2}\right ) \,=\,e^3\,,
\label{eq:timescale}
\end{equation}
where $x = \tau_{\mathrm{exp}}\tau_1^{-1}$. For the values
$\tau_1 = 7.8\times 10^{-3}\,$s and $\tau_2 = 9.48\times 10^{-5}\,$s
adopted by NQM07, one obtains 
$\tau_{\mathrm{exp}} \approx 1.30\times 10^{-3}\,$s as the 
corresponding e-folding timescale.

\begin{figure}[!t]
\includegraphics[width=8.5cm]{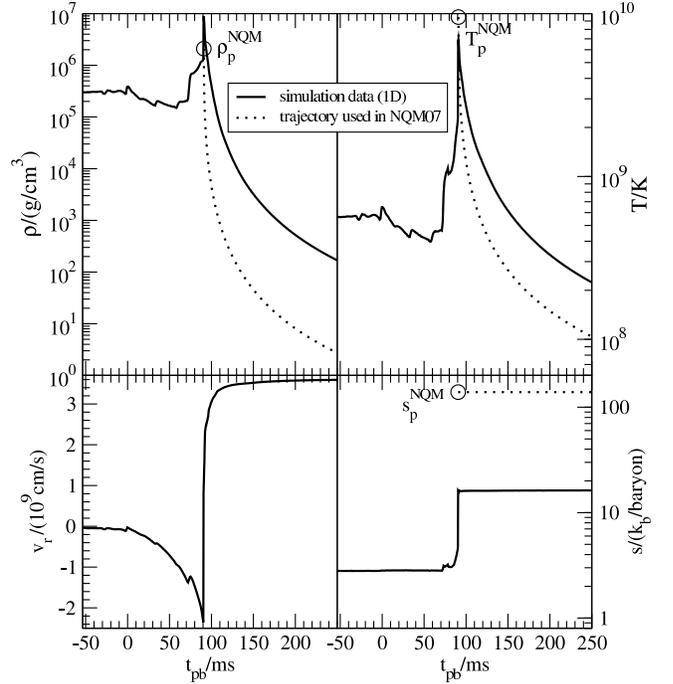}
\caption{
Density (upper left panel), temperature (upper right panel),
radial velocity (lower left panel), and entropy history 
(lower right panel) of a mass shell with
an initial density of $\rho_\mathrm{ini} = 3\times 10^5\,$g$\,$cm$^{-3}$,
which was considered by Ning et al.\ (2007) for their nucleosynthesis
studies. The solid lines correspond to the results of our hydrodynamic 
(1D) simulation, the dotted lines show the expansion behavior of the
shocked shell as described
by the simple analytic model of NQM07 using their parameters. Open
circles indicate the postshock values of density, temperature, and 
entropy assumed by these authors.
}
\label{fig:trho-history}
\end{figure}

Figure~\ref{fig:expansion-time} shows the expansion timescales
measured for the mass shells ejected in our hydrodynamic explosion
models, plotted versus characteristic densities. 
Two e-folding times are given: the red curve corresponds
to the cooling time from $T = 5\times 10^9\,$K to $1/e$ of this value,
the black curve denotes the timescale for the temperature to decrease
from $T = \min(T_{\mathrm{max}},10^{10}\,\mathrm{K})$ to $T/e$, when
$T_{\mathrm{max}}$ is the maximum temperature present in the shell 
before its expansion. The solid lines indicate matter that is swept 
outward by the expanding shock directly, whereas the dashed sections
of the curves belong to matter that was first accreted onto the
forming neutron star before it was later expelled in the
neutrino-driven wind. The black and red lines coincide at low 
preshock densities where the outgoing shock is unable to heat 
the matter to more than $5\times 10^9\,$K, which is also considered
to roughly mark the lower boundary of the regime where NSE can be
established in the shocked gas.
For all preshock densities $\rho \gtrsim 10^3\,$g$\,$cm$^{-3}$,
the expansion timescale is longer than 10$\,$ms, which is at least a 
factor of 10 larger than assumed by NQM07.

In Fig.~\ref{fig:trho-history} the temperature and 
density evolution of a collapsing and ultimately ejected 
mass shell in the C+O layer with an initial density
$\rho_\mathrm{ini} = 3\times 10^5\,$g$\,$cm$^{-3}$ and
an initial temperature $T_\mathrm{ini} = 5.8\times 10^8\,$K
is displayed. NQM07 focused on this shell for their 
nucleosynthesis studies. They assumed that the shell stays at
its initial density and temperature until it is hit by the shock.
This preshock behavior and the conditions in the shocked shell 
considered by NQM07 (dotted lines) are clearly different from
the results of our hydrodynamic model (solid lines).
The mass shell in the simulation lingers near hydrostatic
equilibrium for nearly 100$\,$ms. The
slight density and temperature decrease before the steep rise
does {\em not} signal an expansion of the O-Ne-Mg core: the 
velocity of the considered mass shell is near zero 
until about 60$\,$ms after the start of the simulation
and then becomes increasingly negative (see Fig.~\ref{fig:trho-history},
lower left panel). Instead, the 
differential collapse of the core, i.e.\ the fact that 
the deeper layers start their infall first and contract
faster, leads to a
period of stretching (${\vec \nabla}{\vec v} < 0$) of the
mass shell and therefore to a drop of its density and temperature.
Then the infall of the shell accelerates, triggered by the 
abating pressure support from the already collapsing inner
regions, and the associated compression leads to
a rapid rise of the temperature and density. With the growing 
temperature the carbon and oxygen burning timescales decrease
steeply, but they come close to or become shorter than the 
collapse timescale only when the infall
velocity of the shell is already much larger than the local
sound speed. Therefore the energy release of the nuclear reactions 
decelerates the supersonic collapse of the shell only transiently
and locally,
but does not fundamentally alter its overall dynamics. The next
difference to the analytic NQM07 description is the fact that
the hydrodynamical simulation yields a higher postshock density and 
lower postshock temperature, whose combination corresponds to a
significantly lower postshock entropy (see Fig.~\ref{fig:trho-history}). 
Finally, the decline of 
$T(t)$ and $\rho(t)$ is much slower than considered by NQM07.
This illustrates the different thermodynamical conditions
and expansion behavior of the shock-heated and accelerated
matter. The most relevant differences originate from the 
discrepancy between the shock velocity in the simulations and
the value assumed by NQM07.

\section{Summary and conclusions}
\label{sec:conclusions}

We have shown that the conditions required for a new, 
fast-expansion, modest-entropy r-process scenario 
in the shock-heated ejecta of O-Ne-Mg core supernovae as
proposed recently by NQM07,
are not matched by detailed hydrodynamical explosion models.
From Figs.~\ref{fig:entropy-temperature}--\ref{fig:expansion-time}
it is evident that the expanding mass shells in our 1D as well as
2D simulations never attain the combination of conditions 
identified by NQM07 as favorable for the r-process:
$s_{\mathrm{p}}\sim 150\,k_{\mathrm{B}}$ per nucleon, 
$\tau_{\mathrm{exp}} \sim 1\,$ms, and the postshock temperature
$T_{\mathrm{p}}$ high enough for NSE being established
in the shock-heated matter. Our simulations reveal that either 
the entropy remains too low
(by a factor of 3--5) or the maximum temperature is far from 
that for NSE (approximately by a factor 5). In any case, the 
expansion is roughly ten times slower than needed. The conditions
in explosion models of O-Ne-Mg cores therefore miss those
necessary for the new r-process scenario by at least as much
as current neutrino-wind models fail to produce the conditions
for strong r-processing in the ordinary 
high-entropy wind scenario (where the entropies must typically
be a factor 2--3 larger than provided by the models, see e.g.\
Witti et al.\ 1994, Qian \& Woosley 1996, Thompson et al.\ 2001).

The main reason for the inadequacy of the O-Ne-Mg core explosions
is a significantly slower shock velocity 
(Fig.~\ref{fig:vstoss-density}) than assumed by NQM07,
who took $v_{\mathrm{sh},10} = 1.5$.
Only in progenitor layers with an initial density of less than
about $10^3\,$g$\,$cm$^{-3}$ does the shock reach a speed near
$10^{10}\,$cm$\,$s$^{-1}$ or higher, and therefore the postshock 
entropies exceed 100$\,k_{\mathrm{B}}$ per nucleon 
and the expansion timescale tends to become short. However, the
temperatures of that shock-heated gas then remain so low
that NSE is never achieved. Moreover, these low-density
layers contain two to three orders of magnitude less mass than 
the shells considered by NQM07, and therefore
it is questionable whether they could contribute to the 
r-process inventory of the galaxy on any significant scale,
even if r-process nuclei were able to form there in a way
that does not require a freeze-out from NSE conditions.

We therefore conclude that the newly suggested r-process 
scenario is unlikely to work in supernovae of progenitor
stars with O-Ne-Mg cores. Detailed nucleosynthesis calculations
based on our explosion models confirm this conclusion 
(Hoffman et al.\ 2008). 

The extremely rapid shock acceleration that is necessary
to reach the required high temperatures and entropies and the
very short expansion timescales of core matter in layers
with relatively high initial densities, cannot be obtained 
in present explosion models. In view of the sophistication 
of the 1D and 2D models this failure may point to a 
significant deficit of the progenitor data and assumed
initial conditions. Very rapid rotation, which affects
the structure of the collapsing stellar core already during
the infall stage and shortly after bounce, or very strong 
magnetic fields must be expected to modify the explosion 
conditions compared to our simulations. Such effects would 
require the reinvestigation of the pre-collapse evolution 
of low-mass supernova progenitors in the
8--10$\,M_\odot$ range including rotation and magnetic 
fields. Also the accretion induced collapse (AIC) of rapidly
rotating white dwarfs, which was simulated recently
by Dessart et al.\ (2006, 2007) 
without and with magnetic fields, may deserve a detailed
evaluation of the associated nucleosynthesis. It is,
however, unclear whether AICs occur frequently enough to
be seriously considered as a major site for the production
of high-mass r-process elements, in particular since their
event rate seems to be strongly limited by the potential
massive overproduction of closed neutron shell $N = 50$ 
material (see Dessart et al.\ 2007 and references therein).

\begin{acknowledgements}
We are grateful to K.~Nomoto for providing us with his
progenitor data and to A.~Marek for his contributions to the
microphysics used in the supernova runs. We also thank the 
referee, Raph Hix, for his valuable suggestions to improve
our manuscript. The project was supported by
the Deutsche Forschungsgemeinschaft
through the Transregional Collaborative Research Centers SFB/TR~27
``Neutrinos and Beyond'' and SFB/TR~7 ``Gravitational Wave Astronomy'',
and the Cluster of Excellence EXC~153 ``Origin and Structure of the Universe''
({\tt http://www.universe-cluster.de}). The computations were
done at the High Performance Computing Center Stuttgart (HLRS) under
grant number SuperN/12758.
\end{acknowledgements}

{\small
\bibliographystyle{aa}


}
 
\end{document}